\documentclass[final,3p,times]{elsarticle}

\usepackage{graphicx,setspace}
\usepackage{psfrag}
\usepackage{amsfonts}
\expandafter\let\csname equation*\endcsname\relax
\expandafter\let\csname endequation*\endcsname\relax
\usepackage{amsthm}
\usepackage{mathrsfs}
\usepackage{epstopdf}
\usepackage{float}
\usepackage{color}
\usepackage{subfigure}
\usepackage{tabularx}

\usepackage{dsfont} 
\usepackage{amssymb} 
\usepackage{graphicx,color}
\usepackage{extarrows}
\usepackage{amsmath,amssymb,bm}
\usepackage{booktabs}
\usepackage{graphicx}
\usepackage{array}
\usepackage{tabularx}
\usepackage{caption}
\usepackage{subcaption}
\usepackage{placeins}
\usepackage{xcolor}
\usepackage{microtype}
\usepackage{graphicx}

\usepackage{epstopdf}

\journal{arXiv}

\begin{document}
\newtheorem{definition}{Definition}[section]
\newtheorem{lemma}{Lemma}[section]
\newtheorem{remark}{Remark}[section]
\newtheorem{theorem}{Theorem}[section]
\newtheorem{proposition}{Proposition}
\newtheorem{assumption}{Assumption}
\newtheorem{example}{Example}
\newtheorem{corollary}{Corollary}[section]
\renewcommand{\theequation}{\thesection.\arabic{equation}}
\newcommand{\ajh}{a_{JH}}
\newcommand{\vectm}{\bm m}
\newcommand{\heff}{\bm h_{\mathrm{eff}}}
\newcommand{\R}{\mathbb{R}}
\begin{frontmatter}



\title{Physics-Constrained Neural Flow Maps for Long-Horizon Prediction of Spin Dynamics}


\author{Haoen Feng\fnref{addr1,addr2}}\ead{fenghaoen@stu2025.jnu.edu.cn}
\author{Shenglan Yuan\fnref{addr3}}\ead{shenglanyuan@gbu.edu.cn}
\author{Shirong Lin\corref{cor1}\fnref{addr1,addr4}}\ead{shironglin@gbu.edu.cn}\cortext[cor1]{Corresponding author}

\address[addr1]{\rm School of Physical Sciences, Great Bay University, Dongguan 523000, China}
\address[addr2]{\rm School of Intelligent Systems Science and Engineering, Jinan University, Zhuhai 519000, China}
\address[addr3]{\rm Department of Mathematics, School of Sciences, Great Bay University, Dongguan 523000, China}
\address[addr4]{\rm Great Bay Institute for Advanced Study, Dongguan 523000, China}

\begin{abstract}
Conventional simulation of current-driven magnetization relies on fine-step integration of the spin-transfer-torque Landau--Lifshitz--Gilbert equation, creating a computational bottleneck in parameter sweeps and control searches. In this work, we propose a physics-constrained neural flow map that learns finite-time dynamics directly on the unit sphere. The model maps the current magnetization, spin-torque strength, and requested time span to a future state in a single forward pass. Tangent-space projection and spherical retraction preserve unit magnetization during recursive, composition-consistent rollout. We validate the framework on single-spin trajectories under in-domain torques and previously unseen but stronger drive. Beyond the training horizon, it achieves an in-domain root mean square error of $0.00425$ with norm drift at the $10^{-7}$ level. The flow outperforms an adapted Long Short-Term Memory (LSTM) in in-domain accuracy and geometric stability, although the LSTM retains slightly lower out-of-distribution state error. The resulting geometry-preserving propagator reduces reliance on fine-step integration and enables physically admissible long-horizon prediction.
\end{abstract}


\end{frontmatter}

\section{Introduction}
Current-driven magnetization reversal is a fundamental dynamical process in spintronic devices. At the heart of describing such spin dynamics lies the Landau–Lifshitz–Gilbert (LLG) equation \cite{Lakshmanan2011fascinating}, which governs the time evolution of magnetization. To model the operation of magnetoresistive devices, however, the LLG equation must be extended to include the spin-transfer torque (STT)—a term that conveys angular momentum from a spin-polarized current to the magnetic moment of the free layer\cite{Barashenkov2020Stable,Ghosh2022Unconventional}. Together with field-induced precession and Gilbert damping, the STT critically determines the reversal trajectory, shaping both the switching speed and the stability of the final magnetic state\cite{gilbert2004damping,slonczewski1996current}.
 A quantum-mechanical derivation of an extended LLG equation further relates current-induced torque and damping to spin-wave dynamics and impurity scattering in ferromagnetic wires \cite{edwards2009quantum}. Wen and Xia\cite{wen2017control} studied how current, damping, and magnetic anisotropy affect the reversal time in a single-macrospin model\cite{Soykal2010Strong,Liu2026Tunable,Huang2026Coupling}, providing the spin-transfer-torque Landau--Lifshitz--Gilbert (STT--LLG) control problem adopted here. Device-parameter sweeps, control searches, and uncertainty analyses require the same dynamical equation to be integrated repeatedly over many current settings. Projection-based semi-implicit schemes and second-order methods have improved the stability and efficiency of LLG integration \cite{banas2005numerical,cai2022secondorder}, but the cost of long trajectories is still governed by repeated small-step updates.

Data-driven models offer an alternative way to amortize these repeated integrations. Kovacs \emph{et al}. compressed micromagnetic states with a convolutional autoencoder and predicted LLG dynamics in the latent space \cite{kovacs2019learning}. Exl \emph{et al}. learned field-parameterized magnetization responses through kernel-based dimensionality reduction and low-rank time propagation \cite{exl2020learning,exl2021prediction}. More recently, a latent-manifold approach unified dynamical prediction and control generation within an LSTM representation, demonstrating that simulation and control can be learned in a common framework \cite{zhong2026latent}. Collectively, these studies show that learned propagators can reduce online computational cost, but long autoregressive rollouts, unseen parameters, and physical state constraints continue to delimit their reliable operating range.

The single-macrospin problem does not require further state compression, but it exposes a more direct stability issue. The magnetization direction lies on the unit sphere $\mathbb S^2$, whereas an unconstrained three-dimensional multilayer perceptron output has no reason to satisfy $\|\bm m\|_2=1$. Recent neural modeling of LLG dynamics likewise treats the non-convex unit-norm condition as a central physical constraint and incorporates norm-preserving regularization \cite{dolui2026neural}. Even when one-step errors are small, radial drift and tangential phase error can accumulate through different channels during autoregressive prediction. Radial drift makes the state physically inadmissible; tangential error changes the oscillation phase, damping transient, and asymptotic direction. A long-horizon magnetization surrogate therefore cannot be assessed by one-step mean squared error alone. Closed-loop state error, spherical geometry, and behavior beyond the training parameter boundary must all be examined.

A finite-time flow map provides a natural representation for this purpose. Rather than learning the instantaneous right-hand side of the STT-LLG equation and then invoking a numerical integrator, a flow map directly approximates the evolution operator over a prescribed time span. Neural flow maps have recently been used to reconstruct continuous-time dynamics directly from sparse and irregularly sampled time-series data \cite{xu2025timeseries}. Data-driven evolution maps can obtain multistep forecasts by recursively applying a one-step model \cite{qin2019datadriven}, while parameterized families of equations can be learned by including equation parameters as conditional inputs \cite{qin2021parameterized}. Churchill and Xiu provide an overview and long-horizon benchmarks for this class of flow-map methods \cite{churchill2023flowmap}. Maps at different time spans can also be coupled through composition or semigroup consistency \cite{chen2023deeposg}. Large-step Hamiltonian flow-map models further illustrate the value of learning finite-time evolution directly \cite{ripken2026hfm}. Hamiltonian conservation, however, does not transfer directly to STT-LLG dynamics with damping and nonconservative torque. For magnetization dynamics, the spherical geometry must be built into the neural update itself.

To address these limitations, we propose a parameter-conditioned, physics-constrained neural flow map. The model directly maps the current magnetization, normalized spin-transfer-torque strength, and requested time span to a future magnetization state. A residual decoder predicts a tangential increment that is retracted to the unit sphere, while multi-span training and composition consistency allow the same network to operate across temporal scales. We validate the framework on the single-macrospin problem under both in-domain torques and unseen stronger drive, with all future-window predictions generated fully autoregressively. Across five independent runs, the model attains an average in-domain root mean square error (RMSE) of $4.25\times10^{-3}$ and maintains norm drift at the $10^{-7}$ level. An adapted LSTM baseline is less accurate and substantially less stable geometrically in-domain, although it achieves slightly lower state error under stronger-drive conditions during out-of-distribution (OOD) evaluation. Ablations identify spherical retraction as the source of radial stability and composition consistency as an aid to closed-loop accuracy. The resulting framework replaces repeated fine-step integration with a reusable, geometry-preserving finite-time propagator.

The remainder of this paper is organized as follows. Section 2 introduces the STT-LLG problem and the reference dynamics. Section 3 presents the physics-constrained neural flow map. Section 4 describes the training and evaluation protocol. Section 5 reports the validation, generalization, and ablation results. Section 6 concludes the paper and discusses the remaining challenges.

\section{Physical Problem and Reference Dynamics}

\subsection{Single-macrospin STT--LLG model}

Let $\vectm=(m_x,m_y,m_z)^\mathsf{T}\in\mathbb S^2$ denote the magnetization direction. Time $t$ is expressed in nanoseconds throughout the numerical implementation. We solve
\begin{equation}
\frac{\mathrm d\vectm}{\mathrm dt}
=\frac{1}{1+\alpha^2}\left[
-\vectm\times\heff
-\alpha\vectm\times(\vectm\times\heff)
-\ajh\vectm\times(\vectm\times\bm n)
+\alpha\ajh\vectm\times\bm n
\right],
\label{eq:sttllg}
\end{equation}
where
\begin{equation}
\heff=(h_xm_x,0,h_zm_z)^\mathsf{T},\qquad
\vectm(0)=\bm n=\left(\sin\frac{\pi}{4},0,\cos\frac{\pi}{4}\right)^\mathsf{T}.
\end{equation}
We set $h_x=0.02$, $h_z=0.5$, and $\alpha=0.02$. The torque structure in Eq.~\eqref{eq:sttllg} follows the macrospin model \cite{wen2017control}. The fine reference step is $\Delta t=0.045\,\mathrm{ns}$. During closed-loop evaluation, one call to the neural flow map spans ten fine steps and therefore advances the state by $0.45\,\mathrm{ns}$.

The coefficient $\ajh$ is not the electric current itself. It is the dimensionless spin-transfer-torque strength normalized by the characteristic field $H_0$. In the macrospin model \cite{wen2017control},
\begin{equation}
\ajh=\frac{a_J}{H_0}
=\frac{\hbar\eta I}{2e\,\mu_0 S d_{\mathrm F} M_s H_0},
\label{eq:ajh}
\end{equation}
where $I$ is the injected current, $\eta$ is the electron spin-polarization ratio, $e$ is the elementary charge, $\mu_0$ is the vacuum permeability, and $H_0$ is the characteristic field. The quantities $S$, $d_{\mathrm F}$, and $M_s$ denote the free-layer area, thickness, and saturation magnetization, respectively. Thus, $|\ajh|$ measures the strength of the current-induced torque relative to the magnetic-field scale, while its sign encodes the current polarity and its action relative to the pinned-layer magnetization. We adopt the convention $\ajh<0$. The training interval $|\ajh|=0.03$--$0.08$ spans relatively weak to strong STT drive, whereas the out-of-domain interval $|\ajh|=0.08001$--$0.10$ represents unseen, stronger drive.

The first two terms on the right-hand side of Eq.~\eqref{eq:sttllg} describe field-induced precession and damping; the last two combine damping-like and field-like current-driven contributions. Under the sign convention used here, $\ajh$ is negative and the pinned-layer direction $\bm n$ is also the initial magnetization direction. The unit sphere is not inferred from additional observations but is the state space of the continuous model: whenever $\|\vectm\|_2=1$, every cross-product term on the right-hand side is orthogonal to $\vectm$. A discrete scheme or neural surrogate that does not explicitly respect this structure can still leave the sphere because of finite-precision and fitting errors.

\subsection{Discrete reference flow}

Reference trajectories are generated by a four-stage explicit propagator followed by normalization to the unit sphere after every full step. The effective field $\heff(\vectm_k)$ is computed once at the beginning of the step and reused across all stages. The resulting update is therefore a frozen-effective-field discrete flow with post-step projection. This scheme differs from standard fourth-order Runge-Kutta (RK4) method , which recomputes the complete right-hand side at every intermediate stage. We treat the four-stage procedure as the data-generation rule and do not claim fourth-order accuracy for the underlying continuous equation. The post-step projection is consistent with the spherical geometry of magnetization states \cite{lewis2003geometric,hairer2006geometric}. Specialized treatments of fixed spin length and numerical stability in LLG integration have also been studied extensively \cite{mentink2010stable}, directly motivating the explicit norm preservation in our neural update.

More precisely, one fine step first fixes $\bm h_{\mathrm{eff},k}=\heff(\vectm_k)$, advances the remaining cross-product structure through four stages, combines the stage increments, and then applies $\vectm_{k+1}\leftarrow\vectm_{k+1}/\|\vectm_{k+1}\|_2$. The pregenerated data and the online targets used during training follow this identical propagation path and coincide elementwise within each training batch. The model learns this discrete reference flow, and every RMSE reported below is defined relative to it.

\section{Physics-Constrained Neural Flow Map}

\subsection{Parameter-conditioned finite-time map}

For constant $\ajh$, Eq.~\eqref{eq:sttllg} is an autonomous first-order system whose finite-time evolution is determined by the current state and parameter. We learn
\begin{equation}
\widehat{\vectm}_{k+\delta}
=F_\theta(\vectm_k,\ajh,\Delta t)
\end{equation}
rather than regressing the complete curve $\vectm(t)$ at once. The input vector is
\[
\bm x=\left[m_x,m_y,m_z,\bar a_{JH},d,d^2,\sin(\pi d),\cos(\pi d)\right],
\]
where $\bar a_{JH}$ is standardized using training-set statistics only and $d=\Delta t/\Delta t_{\max}=\delta/10$. During training, $\delta$ is sampled uniformly from $1,\ldots,10$, so $\Delta t\in[0.045,0.45]\,\mathrm{ns}$. Closed-loop evaluation always uses $\delta=10$.

\begin{figure}[htbp]
  \centering
  \includegraphics[width=\textwidth]{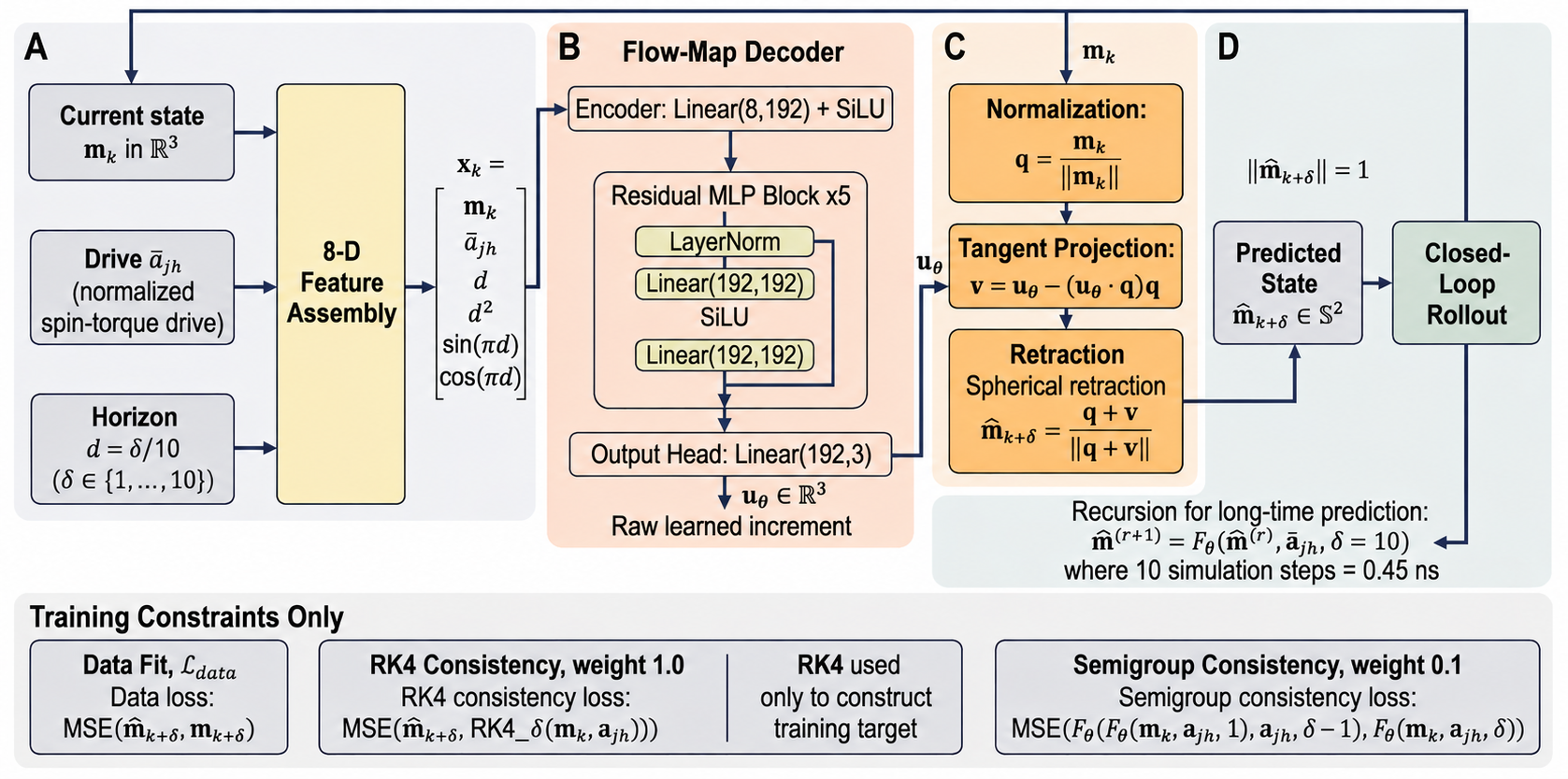}
  \caption{Physics-constrained neural flow map. The current magnetization, normalized spin-transfer-torque strength, and requested prediction span are encoded as eight input features. A residual  multi-layer perceptron produces a three-dimensional increment; tangent-space projection and spherical retraction constrain the predicted state to $\mathbb S^2$, after which the prediction is fed back for closed-loop propagation. Training combines data fitting, duplicate supervision equivalent to the data term, and single-split composition consistency. Each closed-loop call spans ten fine steps, corresponding to $0.45\,\mathrm{ns}$.}
  \label{fig:architecture}
\end{figure}

An $8\to192$ linear layer with SiLU activation encodes the input, followed by five residual blocks of width 192. Each residual block contains LayerNorm, two linear transformations, and SiLU activations. The final layer produces a raw three-dimensional increment $\bm u$. The architecture has 374,787 trainable parameters. Figure~\ref{fig:architecture} summarizes the data path, geometric layer, and training-only constraints.

The span features $d,d^2,\sin(\pi d),\cos(\pi d)$ are not additional physical states but a multiscale encoding of the requested finite-time interval. The same $\vectm$ and $\ajh$ can correspond to different $\Delta t$, so the network must distinguish the family of maps being learned. The current parameter is standardized only by the training-set mean and standard deviation; validation, testing, and OOD evaluation all reuse the same transformation, ensuring that out-of-range parameters remain out of range in standardized coordinates.

\subsection{Spherical update and training objective}

Let $\bm q=\vectm_k/\|\vectm_k\|_2$ be the normalized base point. The raw network output is first projected onto $T_{\bm q}\mathbb S^2$:
\begin{equation}
\bm v=\bm u-(\bm u\cdot\bm q)\bm q,
\qquad
F_\theta(\vectm_k,\ajh,\Delta t)
=\frac{\bm q+\bm v}{\|\bm q+\bm v\|_2}.
\label{eq:retraction}
\end{equation}
Provided the denominator is nonzero, Eq.~\eqref{eq:retraction} enforces unit norm by construction. The retraction does not guarantee small state error. It closes only the radial-drift channel; tangential phase and asymptotic-direction errors must still be learned from the data and dynamical constraints.

The tangent projection removes the component of $\bm u$ parallel to $\bm q$, aligning the locally learned degrees of freedom with the two-dimensional tangent plane of the sphere. Normalized retraction then maps the finite increment back to the sphere. The network is therefore not required to discover $\|\vectm\|_2=1$ during training, nor must a penalty coefficient trade state accuracy against norm preservation. Retraction is one local geometric choice among others. An exponential map also preserves norm, but the two maps produce different states for finite increments, so their accuracy must be compared experimentally.

The implemented loss and its equivalent form are
\begin{equation}
\mathcal L=\mathcal L_{\mathrm{data}}+1.0\,\mathcal L_{\mathrm{dup}}
+0.1\,\mathcal L_{\mathrm{comp}}
=2.0\,\mathcal L_{\mathrm{data}}+0.1\,\mathcal L_{\mathrm{comp}},
\label{eq:loss}
\end{equation}
where the pregenerated trajectory label is denoted by $\vectm^{\mathrm{data}}_{k+\delta}$. The label obtained by calling the same reference propagator online is
\begin{equation*}
\vectm^{\mathrm{on}}_{k+\delta}=R_\delta(\vectm_k,\ajh).
\end{equation*}
The two supervised terms are
\begin{align}
\mathcal L_{\mathrm{data}}&=\operatorname{MSE}\!\left(F_{\theta,\delta}(\vectm_k,\ajh),\vectm^{\mathrm{data}}_{k+\delta}\right),\\
\mathcal L_{\mathrm{dup}}&=\operatorname{MSE}\!\left(F_{\theta,\delta}(\vectm_k,\ajh),\vectm^{\mathrm{on}}_{k+\delta}\right).
\end{align}
Both labels use the same initial state, parameter, span, and frozen-effective-field propagation path. They are identical elementwise in the actual training batches, so $\mathcal L_{\mathrm{dup}}\equiv\mathcal L_{\mathrm{data}}$. The implementation calls $\mathcal L_{\mathrm{dup}}$ the ``RK4 loss''. Because it uses the same target as the data term, we interpret it as supervised reweighting rather than an independent source of physical consistency. The composition term imposes
\begin{equation}
F_{\theta,\delta-1}\!\left(F_{\theta,1}(\vectm,\ajh),\ajh\right)
\approx F_{\theta,\delta}(\vectm,\ajh).
\end{equation}
This relation applies only to sampled spans with $\delta\ge2$. Samples with $\delta=1$ do not contribute to the composition term, and gradients pass through both the direct and nested branches. The condition is enforced only within an autonomous, constant-$\ajh$ parameter slice. It is an empirical ``one step plus the remaining steps'' decomposition, not a proof of a strict semigroup property under arbitrary time partitions. Its motivation is consistent with composition learning for variable-span evolution operators \cite{chen2023deeposg}; the use of multistep training to improve recursive stability also follows prior flow-map and magnetization-surrogate studies \cite{churchill2023flowmap,exl2021prediction}.

The data term directly supervises the pregenerated trajectories, the duplicate term applies the same target a second time, and the composition term compares direct and decomposed predictions without introducing a new label. Consequently, changing the so-called ``RK4 weight'' in Table~\ref{tab:ablation} simply changes the total coefficient on the data error from 2.0 to 1.5 or 1.0, along with its ratio to the composition term. The weights 2.0 and 0.1 define the present main-model configuration; we do not interpret them as universally optimal values derived from theory.

Training uses AdamW with an initial learning rate of $8\times10^{-4}$, weight decay of $10^{-6}$, cosine annealing to $4\times10^{-5}$, a batch size of 4096, gradient-norm clipping at 1.0, and at most 8000 iterations.

\section{Training and Evaluation Protocol}

\subsection{Parameter regimes and model selection}

The in-domain parameter grid is $\ajh\in[-0.08,-0.03]$, corresponding to relatively weak through strong STT drive with $|\ajh|=0.03$--$0.08$. Parameter points are partitioned into training, validation, and test sets in a $10{:}1{:}1$ ratio. Complete parameter trajectories, rather than individual time slices, are the units of partitioning, so time slices from the same parameter never cross subsets. A separate OOD set with $\ajh\in[-0.10,-0.08001]$ tests unseen but stronger drive and is strictly separated from the in-domain boundary.

Checkpoints are selected exclusively by full closed-loop RMSE over the future validation window. Neither the test set nor the OOD set participates in model selection; both are evaluated only after the checkpoint has been fixed. The main results aggregate five independent training runs with the same data partition and protocol but independent network initialization and minibatch sampling.

\subsection{Closed-loop rollout and metrics}

Training trajectories cover only $0\leq t\leq99.99\,\mathrm{ns}$. At evaluation time, the model reads the reference state once at $t=99.99\,\mathrm{ns}$ and then receives neither future ground truth nor further calls to the reference integrator. It advances for 222 closed-loop neural steps of $\Delta t=0.45\,\mathrm{ns}$, ending at $199.89\,\mathrm{ns}$. In-domain testing measures future-window rollout within the observed drive-strength range. The stronger-drive OOD setting combines future-window propagation with one-sided parameter extrapolation. Test and OOD parameters are never used for checkpoint selection, although the validation set uses the same future time window; the temporal window itself is therefore not fully blinded.

This protocol differs from teacher forcing and sliding restarts. Every future prediction is fed back as the next network input, allowing errors to propagate into all subsequent states; no reference state after the boundary is used for correction. Each evaluation call uses the largest locally observed training span, $0.45\,\mathrm{ns}$. Extrapolation occurs in the number of successive compositions and in cumulative absolute time, not in the span of an individual call. Because the autonomous system does not explicitly depend on absolute time, and because we do not quantify whether state distributions before and after the boundary overlap, we describe this setting as a ``closed-loop rollout beyond the observed trajectory horizon'' rather than asserting that the post-boundary state distribution is necessarily OOD.

The componentwise RMSE is
\begin{equation}
\mathrm{RMSE}=\sqrt{\frac{1}{3NT}\sum_{i=1}^{N}\sum_{j=1}^{T}
\|\widehat{\vectm}_{ij}-\vectm_{ij}^{\mathrm{ref}}\|_2^2},
\end{equation}
where $N$ is the number of trajectories and $T$ is the number of extrapolated time points. Geometric error is measured by the maximum norm drift,
$\max_{i,j}|\|\widehat{\vectm}_{ij}\|_2-1|$. Means and sample standard deviations are reported over five independent runs. Single-trajectory and ablation results are excluded from these aggregate statistics.

\subsection{Adapted LSTM baseline}
To compare the inductive biases of a finite-time flow map and recurrent latent dynamics, we construct an LSTM baseline from the architecture of Zhong \emph{et al}.\cite{zhong2026latent} and the associated public implementation. The original method jointly learns quantum dynamics and control and therefore cannot be transferred line by line to STT--LLG. We retain the reported three-layer LSTM with hidden dimension 256, dropout 0.1, Adam with learning rate $10^{-3}$, and batch size 128, together with scheduled teacher forcing and chunked rollouts from the public implementation. Quantum-specific inputs are replaced by $[m_x,m_y,m_z,\bar a_{JH}]$, the dynamical decoder is modified to output the three magnetization components, and the task-specific control decoder and control losses are removed. The adapted LSTM has 1,321,731 trainable parameters, compared with 374,787 for the Flow model.

The LSTM and Flow models use identical parameter-trajectory splits, training windows, and future evaluation windows, with five independent runs for each method. Checkpoints for both are selected solely by full closed-loop RMSE on validation parameters over $99.99$--$199.89\,\mathrm{ns}$. The LSTM reads the observed sequence from $0$ to $99.99\,\mathrm{ns}$ at $0.45\,\mathrm{ns}$ intervals to warm up its hidden state, then generates the next 222 steps fully autoregressively. Flow reads only the current state at the boundary. Neither method receives ground truth or calls the reference integrator within the future window. The experiment compares complete surrogate methods under the same data split and deployment horizon; it is not a parameter-matched, information-matched, or module-by-module controlled comparison.

\section{Model Validation and Generalization}

\subsection{Long-horizon dynamical prediction}

Table~\ref{tab:main-results} summarizes future-window performance over five independent training runs. The in-domain test RMSE is $0.00425\pm0.00126$. When the normalized spin-transfer-torque strength is increased beyond the training range, the RMSE rises to $0.0259\pm0.00656$, approximately 6.1 times the in-domain value. The stronger drive clearly makes dynamical extrapolation more difficult, but every model completes all 222 closed-loop steps without numerical divergence.

\begin{table}[htbp]
\centering
\caption{Long-horizon closed-loop performance of the Flow model over five independent training runs. RMSE is reported as mean $\pm$ sample standard deviation. Maximum norm drift is the maximum over the complete evaluation set and time window.}
\label{tab:main-results}
\begin{tabular}{lccc}
\toprule
Evaluation regime & $\ajh$ range & RMSE & Maximum norm drift \\
\midrule
In-domain & $[-0.08,-0.03]$ & $0.00425\pm0.00126$ & $1.788\times10^{-7}$ \\
OOD & $[-0.10,-0.08001]$ & $0.0259\pm0.00656$ & $1.192\times10^{-7}$ \\
\bottomrule
\end{tabular}
\end{table}

Unlike the state error, the geometric error is nearly invariant to the parameter regime and training randomness. Maximum norm drift remains at the scale of single-precision roundoff in both the in-domain and OOD evaluations, confirming that spherical retraction remains active at every closed-loop update. This result guarantees only that each predicted state lies on the admissible manifold; tangential phase, damping rate, and asymptotic direction remain determined by the learned dynamics.

\subsection{Comparison with the adapted LSTM baseline}

\begin{figure}[htbp]
  \centering
  \includegraphics[width=\textwidth,trim=0 0 0 40,clip]{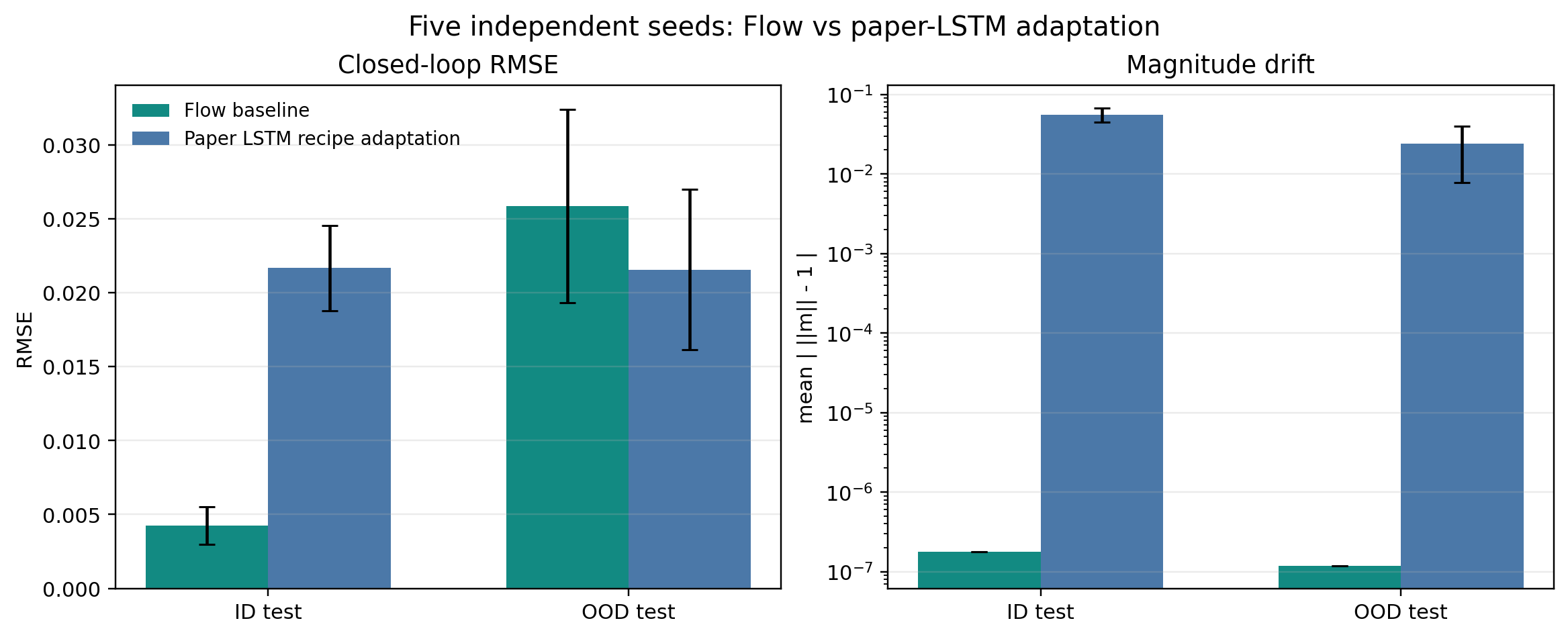}
  \caption{Closed-loop comparison between Flow and the adapted LSTM over five independent training runs. The left panel shows componentwise RMSE for in-domain testing and stronger-drive OOD evaluation; the right panel shows maximum norm drift on a logarithmic scale. Bars and error bars denote the mean and sample standard deviation over five runs. Both methods use the same parameter split and future window, but LSTM warms up its hidden state with the complete observed history, whereas Flow reads only the boundary state.}
  \label{fig:flow-lstm}
\end{figure}

Figure~\ref{fig:flow-lstm} compares five independent runs under the same data split and future window. On the in-domain test set, Flow achieves $0.004247\pm0.001258$ RMSE, compared with $0.021673\pm0.002875$ for LSTM. Flow therefore reduces the error by approximately 80.4\%, a 5.1-fold difference. Under stronger-drive OOD evaluation, Flow and LSTM obtain RMSE values of $0.025859\pm0.006555$ and $0.021570\pm0.005417$, respectively, giving LSTM approximately 16.6\% lower state error. The in-domain advantage therefore does not imply a uniform ranking over all parameter regimes.

The contrast in geometric stability is more decisive. Mean runwise maximum norm drift for Flow is $1.788\times10^{-7}$ in-domain and $1.192\times10^{-7}$ OOD. The corresponding LSTM values are $(5.577\pm1.143)\times10^{-2}$ and $(2.391\pm1.606)\times10^{-2}$, a difference of more than five orders of magnitude. The recurrent hidden state allows LSTM to exploit the complete observation history and to achieve slightly lower tangential state error in the OOD interval, but its three-dimensional output has no unit-sphere guarantee. Flow propagates from the current state alone, yet its explicit geometric layer removes radial error from closed-loop propagation.

\FloatBarrier

Figure~\ref{fig:flow-lstm-waveform} gives a single-trajectory waveform comparison at the in-domain boundary $\ajh=-0.03$ to examine how temporal training coverage affects closed-loop phase error. During evaluation, every model reads reference information only through $49.995\,\mathrm{ns}$ and then extrapolates to approximately $200\,\mathrm{ns}$. The legend labels ``trained to 50'' and ``trained to 100'' indicate that the corresponding training trajectories end at approximately 50 and 100 ns. Extending training coverage to 100 ns reduces the trajectory RMSE from 0.2373 to 0.0332 for Flow and from 0.8291 to 0.1445 for LSTM, showing that both autoregressive surrogates benefit from longer training trajectories. Under both training windows, the oscillation phase and decay envelope predicted by Flow remain closer to the reference flow, and Flow preserves unit norm throughout; LSTM's waveform deviation is accompanied by visible norm drift. This representative temporal-coverage experiment is not included in the five-run statistics of Fig.~\ref{fig:flow-lstm}.

\begin{center}
\begin{minipage}{0.94\textwidth}
  \centering
  \includegraphics[width=\linewidth,trim=0 0 0 55,clip]{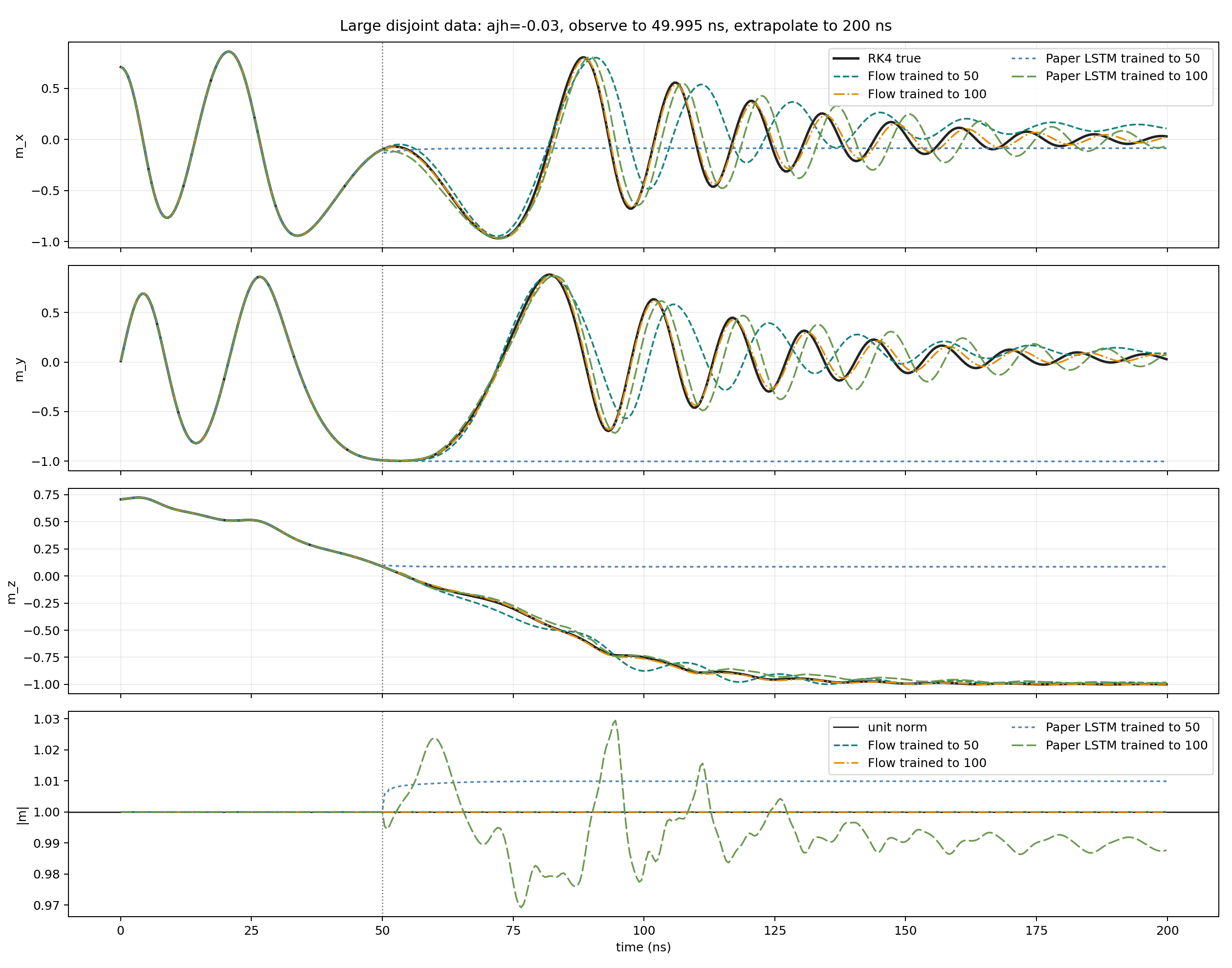}
  \captionsetup{hypcap=false}
  \captionof{figure}{Flow--LSTM waveform comparison under different temporal training horizons. The black curve is the discrete reference flow at $\ajh=-0.03$. Green and orange denote Flow trained to approximately 50 and 100 ns; blue and light green denote the corresponding adapted LSTM models. All methods begin closed-loop extrapolation at $t=49.995\,\mathrm{ns}$. The first three panels show the magnetization components, and the bottom panel shows the norm.}
  \label{fig:flow-lstm-waveform}
\end{minipage}
\end{center}
\FloatBarrier
\subsection{Generalization to unseen stronger drive}

\begin{figure}[htbp]
  \centering
  \includegraphics[width=\textwidth,trim=0 0 0 18,clip]{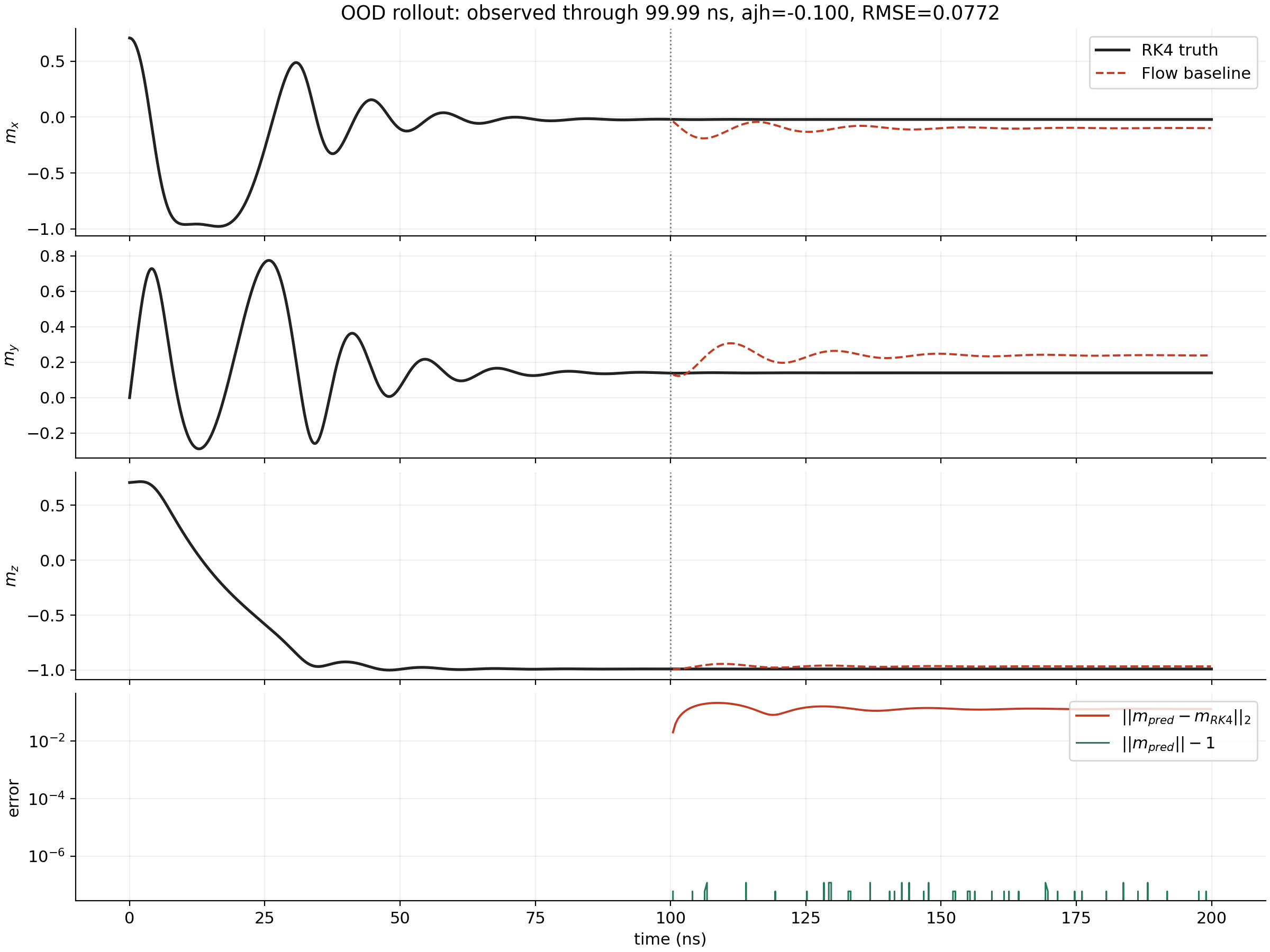}
  \caption{Joint temporal and parameter extrapolation under the unseen, stronger drive $\ajh=-0.10$. Closed-loop propagation begins at $t=99.99\,\mathrm{ns}$ and preserves unit norm throughout. Relative to the reference flow, the predicted asymptotic direction is systematically shifted in the $m_x$ and $m_y$ components.}
  \label{fig:rollout-ood}
\end{figure}

Parameter extrapolation changes the character of the error. Figure~\ref{fig:rollout-ood} shows the unseen parameter $\ajh=-0.10$. Early in the closed-loop rollout, the model still reproduces the decay trend of the reference trajectory, but it subsequently converges to a shifted asymptotic direction in $m_x$ and $m_y$. The representative trajectory has RMSE 0.0772, while $\|\widehat{\vectm}\|_2-1$ remains at the $10^{-7}$ level. This contrast directly separates geometric stability from dynamical accuracy: the spherical constraint prevents radial drift but cannot remove tangential error caused by parameter extrapolation.

\subsection{Ablation of geometric and composition constraints}

The ablation study addresses two distinct questions: whether the spherical update controls geometric stability and whether composition consistency improves closed-loop accuracy. All variants form a single paired experiment with shared initialization and minibatch order, isolating the structural change. The full model obtains in-domain and OOD RMSE values of 0.006599 and 0.030187. Removing composition consistency increases them to 0.007005 and 0.031224. Reducing or removing the duplicate data term also changes the error, but these variants simply reduce the total weight on data supervision from 2.0 to 1.5 or 1.0.

\begin{table}[htbp]
\centering
\caption{Controlled single-factor ablations. All variants share initialization and minibatch order; each result is from one paired run.}
\label{tab:ablation}
\footnotesize
\begin{tabularx}{\textwidth}{>{\raggedright\arraybackslash}Xccc}
\toprule
Model variant & \shortstack{In-domain\\RMSE} & \shortstack{OOD\\RMSE} & \shortstack{In-domain maximum\\norm drift} \\
\midrule
Full model: total data weight 2.0 + comp(0.1) + retraction
  & 0.006599 & 0.030187 & $1.788\times10^{-7}$ \\
Remove composition
  & 0.007005 & 0.031224 & $1.788\times10^{-7}$ \\
Duplicate-term weight $1.0\rightarrow0.5$ (total data weight $2.0\rightarrow1.5$)
  & 0.006817 & 0.030482 & $1.788\times10^{-7}$ \\
Remove duplicate term (total data weight $2.0\rightarrow1.0$)
  & 0.007406 & 0.033067 & $1.788\times10^{-7}$ \\
Retraction $\rightarrow$ exponential map
  & 0.007539 & 0.036309 & $1.192\times10^{-7}$ \\
Retraction $\rightarrow$ raw output
  & 0.007321 & 0.036993 & $5.065\times10^{-3}$ \\
\bottomrule
\end{tabularx}
\end{table}

The raw-output variant reaches an in-domain maximum norm drift of $5.065\times10^{-3}$, approximately $2.83\times10^4$ times that of the full model, while its OOD RMSE increases to 0.036993. The exponential map likewise preserves unit norm but yields higher closed-loop RMSE than spherical retraction. Geometric admissibility is therefore necessary but does not determine the accuracy of a finite-step update. Differences among the composition and data-weight variants are smaller than the cross-seed variation in the main experiment and should be interpreted only as evidence from a controlled single run. By contrast, the effect of spherical retraction on radial stability spans four orders of magnitude.

\section{Conclusion and Challenges}\

We have proposed a physics-constrained neural flow map that directly maps the current magnetization, spin-torque strength, and requested time span to a future state on the unit sphere. The same finite-time propagator is applied recursively for long-horizon prediction without access to future reference states. Validation on the macrospin problem shows accurate in-domain dynamics and near-machine-precision norm preservation under both in-domain and unseen stronger drives. Compared with the adapted LSTM, Flow is substantially more accurate and geometrically stable in-domain; the LSTM's slightly lower OOD state error identifies parameter extrapolation as the main remaining challenge rather than geometric stability.

By converting repeated fine-step integration into a reusable, geometry-preserving evolution operator, the framework provides a physically admissible surrogate for long-horizon magnetization dynamics. Future work will improve torque-parameter generalization, extend the model to multiple initial states and time-dependent currents, and compare alternative learned propagators under matched information and computational budgets.

The central modeling choice is to learn finite-time evolution directly rather than the instantaneous right-hand side or an entire time curve. For constant $\ajh$, the current magnetization is already a Markov state, so the network requires no additional history. Span encoding allows one model to represent evolution over 1--10 fine steps, and the longest span is applied recursively in closed loop. The composition loss further aligns a direct long-span prediction with a decomposition into one step plus the remaining steps. This design reflects the compositional structure of evolution operators \cite{chen2023deeposg} and matches the way the model is ultimately deployed for long-horizon prediction.

The LSTM baseline provides a complementary comparison. Its hidden state aggregates the complete observation history and yields slightly lower RMSE under unseen, stronger drive, but the known spherical state space is absent from its output parameterization. Flow carries no hidden history, is more accurate in-domain, and maintains near-machine-precision norm constraints in both parameter regimes. The waveform comparison also shows that extending training trajectories from approximately 50 to 100 ns improves both surrogates without changing Flow's phase and norm advantage on the representative in-domain trajectory. These results do not support the blanket claim that Flow is superior under every distribution. Instead, the two inductive biases act on different error channels: recurrent memory can improve some aspects of parameter extrapolation, whereas explicit geometry removes radial drift directly. State accuracy and physical admissibility must therefore be reported together in magnetization-control applications.

The experiments expose two error channels in long-horizon magnetization surrogates. Tangent projection and spherical retraction almost completely eliminate radial drift; removing the geometric layer increases the norm error by four orders of magnitude. Tangential error, however, can still shift the oscillation phase and asymptotic direction, particularly under unseen, stronger STT drive. The trajectory in Fig.~\ref{fig:rollout-ood} remains on the sphere but converges to a displaced direction. Physical constraints are therefore not a substitute for accuracy. They exclude inadmissible errors from the model output space, allowing the remaining error to be interpreted in dynamical terms.

Results under parameter extrapolation are consistent with limits reported in earlier magnetization-surrogate and flow-map studies: unseen parameters make reduced propagation more difficult, and long recursion beyond the training horizon can amplify local prediction error \cite{kovacs2019learning,exl2021prediction,churchill2023flowmap}. Our results further show that parameter extrapolation does not automatically violate an explicit geometric constraint, even when tangential state error grows substantially. This separation points to concrete improvements, including denser sampling near parameter boundaries, parameter-adaptive spans, and conditional representations that better resolve changes in asymptotic direction. Joint learning of dynamics and control on a latent manifold has shown promising scalability in quantum control \cite{zhong2026latent}; extending that idea to magnetization control will require both spherical structure and time-dependent drive information.

The principal weakness of the present Flow model is that its dynamical extrapolation is considerably less robust than its geometric preservation. Once the normalized torque strength leaves the training range, mean RMSE rises from 0.00425 in-domain to 0.0259 OOD. The representative trajectory remains exactly on the unit sphere but converges to a shifted asymptotic direction in $m_x$ and $m_y$. Spherical retraction therefore eliminates radial drift, but the current parameter conditioning does not fully capture phase evolution and asymptotic-state migration under unseen drive. The LSTM's slightly lower state error in the same OOD interval provides further evidence that the present Flow representation can be improved outside the training distribution.

Long-horizon accuracy also depends strongly on the temporal coverage of the training trajectories. In Fig.~\ref{fig:flow-lstm-waveform}, extending the Flow training horizon from approximately 50 to 100 ns reduces the representative-trajectory RMSE from 0.2373 to 0.0332. Multi-span supervision and composition consistency have therefore not eliminated phase accumulation caused by insufficient temporal coverage. Future work should prioritize sampling near parameter boundaries and over long transients, together with conditional representations that describe drive-dependent asymptotic states or parameter-adaptive spans. The present flow map is learned only for a fixed initial state and constant current. Practical magnetization control will require coverage of multiple initial states and time-dependent drives, followed by stability tests under closed-loop control sequences. Looking forward, this work not only provides a reliable and efficient computational framework for magnetization dynamics, but also lays a solid foundation for broader investigations of spin systems. The methodology can be naturally extended to atomistic spin dynamics and micromagnetics, opening new avenues for exploring a wide spectrum of magnetic phenomena in both existing and emerging spintronic systems\cite{Lakshmanan2008Dynamic,Beula2010Nonlinear,Birch2021Topological,Henner2026Quantum,Zhang2026Skyrmion}.

\bigskip
\noindent\textbf{Data availability}

The numerical algorithms and source code that support the findings of this study are available from the corresponding author upon reasonable request.

\bigskip
\noindent\textbf{Declaration of competing interest}

No author associated with this paper has disclosed any potential or pertinent conflicts which
may be perceived to have impending conflict with this work.

\bigskip
\noindent\textbf{Acknowledgments}

This work was supported by the National Natural Science Foundation of China (Grant No. 12104296), the Guangdong Basic and Applied Basic Research Foundation (Grant No. 2025A1515012560), the Guangdong Introduction Program (Grant No. 2023QN10X753) and National
Foreign Experts Program (Grant No. 111001819820258003).


\begin{thebibliography}{00}

\bibitem{Lakshmanan2011fascinating}
M. Lakshmanan, The fascinating world of the Landau-Lifshitz-Gilbert equation: an overview. \emph{Philosophical Transactions: Mathematical, Physical and Engineering Sciences}, 2011, 369, 1280–1300.

\bibitem{Barashenkov2020Stable}
I. V. Barashenkov, A. Chernyavsky, Stable solitons in a nearly PT-symmetric ferromagnet with spin-transfer torque. \emph{Physica D: Nonlinear Phenomena}, 2020, 409, 132481.

\bibitem{Ghosh2022Unconventional}
S. Ghosh, A. Manchon, J. Železný, Unconventional robust spin-transfer torque in noncollinear antiferromagnetic junctions. \emph{Physical Review Letters}, 2022, 128(9), 097702.


\bibitem{gilbert2004damping}
T.L. Gilbert, A phenomenological theory of damping in ferromagnetic materials. \emph{IEEE transactions on magnetics}, 2004, 40(6): 3443-3449.

\bibitem{slonczewski1996current}
J.C. Slonczewski, Current-driven excitation of magnetic multilayers. \emph{Journal of Magnetism and Magnetic Materials}, 1996, 159(1-2): L1-L7.

\bibitem{edwards2009quantum}
D.M. Edwards, O. Wessely, The quantum-mechanical basis of an extended Landau--Lifshitz--Gilbert equation for a current-carrying ferromagnetic wire. \emph{Journal of Physics: Condensed Matter}, 2009, 21(14): 146002.

\bibitem{wen2017control}
H Wen, J Xia, Control of spins in a nano-sized magnet using electric-current. \emph{Chinese Physics B}, 2017, 26(4): 047501.

\bibitem{Soykal2010Strong}
Ö. O. Soykal, M. E. Flatté, Strong field interactions between a nanomagnet and a photonic cavity. \emph{Physical Review Letters}, 2010, 104(7), 077202.

\bibitem{Liu2026Tunable}
J. Liu, S. Lin, Tunable rotation-associated slow-to-fast light conversion via optomagnonic coupling. \emph{Physical Review A}, 2026, 113(5), 053514.

\bibitem{Huang2026Coupling}
X. Huang, J. Liu, S. Lin, Coupling phase enabled level transitions in pseudo-Hermitian magnon-polariton systems. \emph{Physical Review B}, 2026, 113(17), 174413.

\bibitem{banas2005numerical}
L. Ba\v{n}as, A numerical method for the Landau--Lifshitz equation with magnetostriction. \emph{Mathematical Methods in the Applied Sciences}, 2005, 28(16): 1939-1954.

\bibitem{cai2022secondorder}
Y. Cai, J. Chen, C. Wang, C. Xie, A second-order numerical method for Landau--Lifshitz--Gilbert equation with large damping parameters. \emph{Journal of Computational Physics}, 2022, 451: 110831.

\bibitem{kovacs2019learning}
A. Kovacs, J. Fischbacher, H. Oezelt, M. Gusenbauer, L. Exl, F. Bruckner, D. Suess, T. Schrefl, Learning magnetization dynamics. \emph{Journal of Magnetism and Magnetic Materials}, 2019, 491: 165548.



\bibitem{exl2020learning}
L. Exl, N.J. Mauser, T. Schrefl, D. Suess, Learning time-stepping by nonlinear dimensionality reduction to predict magnetization dynamics. \emph{Communications in Nonlinear Science and Numerical Simulation}, 2020, 84: 105205.

\bibitem{exl2021prediction}
L. Exl, N.J. Mauser, S. Schaffer, T. Schrefl, D. Suess, Prediction of magnetization dynamics in a reduced dimensional feature space setting utilizing a low-rank kernel method. \emph{Journal of Computational Physics}, 2021, 444: 110586.

\bibitem{zhong2026latent}
J.D. Zhong, Z.Y. Ge, F.H. Ren, Z.M. Wang, End-to-end learning of quantum control on latent dynamical manifold. \emph{arXiv:2606.27907}, 2026.

\bibitem{dolui2026neural}
S. Dolui, M. Salman, A neural network framework for modeling the dynamics of the Landau--Lifshitz--Gilbert equation. \emph{Engineering Applications of Artificial Intelligence}, 2026, 181: 115562.

\bibitem{xu2025timeseries}
B. Xu, Z.E. Ho, Y. Huang, Time-series modeling with neural flow maps. \emph{bioRxiv}, 2025.

\bibitem{qin2019datadriven}
T. Qin, K. Wu, D. Xiu, Data driven governing equations approximation using deep neural networks. \emph{Journal of Computational Physics}, 2019, 395: 620-635.



\bibitem{qin2021parameterized}
T. Qin, Z. Chen, J.D. Jakeman, D. Xiu, Deep learning of parameterized equations with applications to uncertainty quantification. \emph{International Journal for Uncertainty Quantification}, 2021, 11(2): 63-82.


\bibitem{churchill2023flowmap}
V. Churchill, D. Xiu, Flow map learning for unknown dynamical systems: Overview, implementation, and benchmarks. \emph{Journal of Machine Learning for Modeling and Computing}, 2023, 4(2): 173-201.

\bibitem{chen2023deeposg}
J. Chen, K. Wu, Deep-OSG: Deep learning of operators in semigroup. \emph{Journal of Computational Physics}, 2023, 493: 112498.

\bibitem{ripken2026hfm}
W. Ripken, M. Plainer, G. Lied, T. Frank, O.T. Unke, S. Chmiela, F No\'e, K.R. M\"{u}ller, Learning Hamiltonian flow maps: mean flow consistency for large-timestep molecular dynamics. \emph{arXiv:2601.22123}, 2026.

\bibitem{lewis2003geometric}
D. Lewis, N. Nigam, Geometric integration on spheres and some interesting applications. \emph{Journal of Computational and Applied Mathematics}, 2003, 151(1): 141-170.

\bibitem{hairer2006geometric}
E. Hairer, C. Lubich, G. Wanner, \emph{Geometric numerical integration: structure-preserving algorithms for ordinary differential equations}. Springer, 2006.


\bibitem{mentink2010stable}
J.H. Mentink, M.V. Tretyakov, A. Fasolino, M.I. Katsnelson, T. Rasing, Stable and fast semi-implicit integration of the stochastic Landau-Lifshitz equation. \emph{Journal of Physics: Condensed Matter}, 2010, 22(17): 176001.

\bibitem{Lakshmanan2008Dynamic}
M. Lakshmanan, A. Saxena, Dynamic and static excitations of a classical discrete anisotropic Heisenberg ferromagnetic spin chain. \emph{Physica D: Nonlinear Phenomena}, 2008, 237(7), 885-897.

\bibitem{Beula2010Nonlinear}
J. Beula, M. Daniel, Nonlinear spin excitations in a classical Heisenberg anisotropic helimagnet. \emph{Physica D: Nonlinear Phenomena}, 2010, 239(8), 397-406.

\bibitem{Birch2021Topological}
 M. T. Birch, D. Cortés-Ortuño, N. D. Khanh, S. Seki, A. Štefančič, G. Balakrishnan, Y. Tokura, P. D. Hatton, Topological defect-mediated skyrmion annihilation in three dimensions. \emph{Communications Physics}, 2021, 4(1), 175. 
 
\bibitem{Henner2026Quantum}
V. Henner, A. Nepomnyashchy, T. Belozerova, Quantum vs. classical spin: A comparative study of dipolar spin dynamics and the onset of chaos. \emph{Physica D: Nonlinear Phenomena}, 2026, 493, 135251.

\bibitem{Zhang2026Skyrmion}
Q. Zhang, S. Lin, W. Zhang, Skyrmion generation through the chirality interplay of light and magnetism. \emph{Communications Physics}, 2026, 9(1), 55.
\end{thebibliography}
\end{document}